\documentclass[conference]{IEEEtran}
\IEEEoverridecommandlockouts
\usepackage{cite}
\usepackage{comment}
\usepackage{amsmath,amssymb,amsfonts}
\usepackage{algorithmic}
\usepackage{graphicx}
\usepackage{textcomp}
\usepackage{xcolor}
\usepackage[inkscapelatex=false]{svg}
\usepackage[normalem]{ulem}
\usepackage{booktabs}
\usepackage{colortbl}
\usepackage{multirow}
\usepackage{pdfpages}
\usepackage{subcaption}
\usepackage{balance}
\usepackage{url}
\usepackage{float}
\usepackage{placeins}

\def\BibTeX{{\rm B\kern-.05em{\sc i\kern-.025em b}\kern-.08em
    T\kern-.1667em\lower.7ex\hbox{E}\kern-.125emX}}
\begin{document}

\title{LLM4VV: Evaluating Cutting-Edge LLMs for Generation and Evaluation of Directive-Based Parallel Programming Model Compiler Tests\\
}

\author{\IEEEauthorblockN{Zachariah Sollenberger}
\IEEEauthorblockA{\textit{Computational Research and Programming Lab} \\
\textit{University of Delaware}\\
Newark, DE, USA \\
0009-0008-6669-8778}
\and
\IEEEauthorblockN{Rahul Patel}
\IEEEauthorblockA{\textit{Computational Research and Programming Lab} \\
\textit{University of Delaware}\\
Newark, DE, USA \\
0009-0009-8739-7138}
\and
\IEEEauthorblockN{Saieda Ali Zada}
\IEEEauthorblockA{\textit{Computational Research and Programming Lab} \\
\textit{University of Delaware}\\
Newark, DE, USA \\
0009-0000-5368-4512}
\and
\IEEEauthorblockN{Sunita Chandrasekaran}
\IEEEauthorblockA{\textit{Computational Research and Programming Lab} \\
\textit{University of Delaware}\\
Newark, DE, USA \\
0000-0002-3560-9428}

}

\maketitle

\begin{abstract}
The usage of Large Language Models (LLMs) for software and test development has continued to increase since LLMs were first introduced, but only recently have the expectations of LLMs become more realistic. Verifying the correctness of code generated by LLMs is key to improving their usefulness, but there have been no comprehensive and fully autonomous solutions developed yet. Hallucinations are a major concern when LLMs are applied blindly to problems without taking the time and effort to verify their outputs, and an inability to explain the logical reasoning of LLMs leads to issues with trusting their results. To address these challenges while also aiming to effectively apply LLMs, this paper proposes a dual-LLM system (i.e. a generative LLM and a discriminative LLM) and experiments with the usage of LLMs for the generation of a large volume of compiler tests. We experimented with a number of LLMs possessing varying parameter counts and presented results using ten carefully-chosen metrics that we describe in detail in our narrative. Through our findings, it is evident that LLMs possess the promising potential to generate quality compiler tests and verify them automatically.
\end{abstract}


\section{Introduction}

The process of developing test suites for compilers
that implement programming models is complex, arduous, and costly. 
Unit tests to verify compilers' implementations are usually hand-written by a team of programming experts to ensure test correctness and its compatibility with software specifications that defines the programming models. 
Furthermore, achieving full coverage of all possible combinations in a given set of specifications is an astronomically large task. The complexity of writing unit tests for compilers that implement directive-based programming models is similar - the scope of this paper. 

Recent developments in the training and design of Large Language Models (LLMs) may offer a solution to this complexity, i.e. automate compiler test generation. LLMs are being taught to generate high-quality code, which may allow them to create compiler tests for directive-based parallel programming models based purely on provided specifications. 

However, there are two big issues for automating compiler test generation with LLMs. First, in order for a compiler test to validate the implementation of a given feature, the compiler test itself must correctly expose errors in feature implementations. This is usually confirmed manually by the authors of the compiler test, but if LLMs are used to generate a large volume of compiler tests, then manual confirmation of these tests' correctness becomes incredibly time-consuming and likely unfeasible. Second, LLMs have demonstrated difficulty in generating functional and correct parallel code. This ability would be paramount to the automated creation of parallel code test suites.

To address these issues, we propose a dual-LLM system, consisting of a \textit{discriminative} LLM agent trained to identify valid and invalid compiler tests and a \textit{generative} LLM agent trained to generate compiler tests according to provided specifications. In this dual-LLM system, the generative agent creates a compiler test for a specific combination of directives and clauses, and the discriminative agent evaluates the test and returns it to the generative agent for improvement if the discriminative agent does not believe the test to be valid.

The first step in constructing such a system is choosing existing LLMs to fine-tune for these roles. A reasonable approach for this is to evaluate a selection of LLMs and measure how well they can perform these tasks without any additional training, then isolate the best performers to build on their capabilities later. 

In this paper, we begin building such a dual-LLM system by performing this first step and determining the best candidate LLMs to use as our generative and discriminative agents. We present our findings on how well a small assortment of cutting-edge LLMs were able to perform on both generative and discriminative tasks for directive-based parallel programming compiler tests.

\section{Related Work}
The use of large language models (LLMs) for code generation is becoming increasingly popular in the industry of software engineering. Subsequently, more and more research on the subject is being published to discover new applications and approaches for LLMs in every step of the software development process. Pophale et al.~\cite{pophale2025using} evaluated the efficiency and quality of test generation and generated a suite of compiler tests for the features in OpenMP using ChatGPT-4. Of all the tests that were generated, 86\% of the C tests and 54\% of the Fortran tests correctly compiled. 

Research has also shown that LLM-generated parallel code is often invalid, and human intervention is necessary to ensure compiler test correctness. For example, Nader et al.~\cite{nader2025llm} concluded that Deepseek-R1 had difficulty generating scalable code even though its HPC benchmarks originally suggested a higher quality of code generation. Similarly, Nichols et al.~\cite{nichols2024can} used the ParEval benchmark to evaluate parallel code generation, and the results indicate that LLMs are notably worse at generating parallel code than serial code. 


Chen et al.~\cite{chen2024landscape} explored potential applications for LLMs within HPC. Their paper also analyzed other work that used LLMs to generate code and evaluated efforts to assemble repositories together into organized datasets. Other researchers~\cite{misic2024assessment} evaluated the use of ChatGPT and GitHub Copilot for OpenMP-based code parallelization. Valero et al.~\cite{valero2023comparing} compared the LLMs Llama-2 and GPT-3 in their ability to generate HPC kernels. Similarly, Godoy et al.~\cite{godoy2024large} leveraged GPT-3 to generate high-performance computing (HPC) kernels. They developed a proficiency metric that evaluates the initial 10 responses per prompt, allowing for systematic comparisons. The results reveal that correct C++ outputs are strongly linked to the programming models' adoption and maturity.

Research that explores non-parallel code generation using LLMs can also yield valuable information that contributes to the development of parallel code compiler test generation methods. 

According to Yaacov et al. ~\cite{yaacov2024boosting} human code is more effective than LLM generated code.
Furthermore, Huang et al.~\cite{huang2024generative} demonstrated the benefits and drawbacks of their seven established software generative tasks that were performed by LLMs and pre-trained models.

A very similar project to our own was VALTEST~\cite{taherkhani2024valtest}, which proposed an automated system for detecting the validity of test cases created by LLMs. However, their system would not have worked for our purposes, as it was intended to validate test case generation and not test code generation. Their method also does not review the generated cases themselves, but rather the state of the LLM while it is generating them. This means that their system would not be able to validate human-written code or existing compiler tests. However, another project, LLM4VV~\cite{munley2024llm4vv, sollenberger2024llm4vv} proposes a system for validating compiler test code and evaluates the code itself. In their system, an LLM is given the compiler test and instructed to review it and return its validity. Since this method allows for arbitrary code validation, we experiment with their method and their LLMJ approach in this work.

The above research efforts demonstrate that there are other work that explores the use of LLMs for test code generation, each with their own goals and approaches. Keeping these other projects in mind, we focus on using LLMs to generate tests specifically for directive-based parallel programming models. These widely-used models must be implemented independently by each compiler that supports them, so there is a unique and extensive need for validation tests that cover the full specifications of these models. 

To this end, we chose several LLMs with varying parameter counts and training styles, applied a selection of prompt engineering techniques, and evaluated the generative and discriminative capabilities of these LLMs for test code generation and verification. We intentionally chose LLMs of different sizes so that we could compare how larger and smaller LLMs consume HPC resources, and so that we could measure how LLM size impacts generative capability and discriminative accuracy.


\section{Methodology}

\subsection{Choice of LLMs}\label{AA}
We chose HuggingFace~\cite{huggingfaceHuggingFace}, the Ollama Model Library~\cite{ollamaLibrary}, the Artificial Analysis LLM Leaderboard~\cite{artificialanalysisLeaderboardCompare}, and the Github Leaderboard~\cite{evalplusEvalPlusLeaderboard} as references when selecting the LLMs we would evaluate. We did not consider proprietary LLMs as candidates for our evaluations for two main reasons. 

First, we intend to fine-tune the LLMs that performed well in our evaluations as part of our continued project. Since proprietary LLMs either restrict the methods of fine-tuning to specific services or do not allow fine-tuning at all, they would be unsuitable for our future works regardless of their performance. 

Second, the APIs for proprietary LLMs usually charge by the number of tokens generated. Since our evaluation methods involve large volumes of code generation, they would incur very large usage fees and limit the amount of testing we could conduct. We also did not consider LLMs with more than 85B parameters to be viable candidates. Since these LLMs exhibit a prolonged inference time compared to smaller LLMs, a dual-LLM system built with such models would struggle with extended latency and have reduced performance and usefulness. Higher parameter counts would also increase the cost of running such a system. 

We chose to categorize the LLMs we considered by the approximate number of parameters each LLM contained, since we found that most of the LLMs recommended by our references~\cite{huggingfaceHuggingFace,ollamaLibrary,artificialanalysisLeaderboardCompare,evalplusEvalPlusLeaderboard} had parameter counts very close to 7 billion, 32 billion, or 70 billion. We then selected five high-ranking and well-reviewed LLMs for each category, yielding a total of 15 LLMs. Then, after a secondary cross-comparison of our selected LLMs, the two best LLMs from each category were retained for evaluation. Finally, to simplify our nomenclature, the categories were renamed from "7B", "32B", and "70B" to "Small", "Medium", and "Large", respectively.

\begin{table}[ht]
    \centering
    \caption{Preliminary LLMs Considered for Evaluation}
    \resizebox{0.48\textwidth}{!}{%
    \begin{tabular}{|l|l|l|}
        \hline
        \textbf{Small} & \textbf{Medium} & \textbf{Large} \\
        \hline
        Gemma-7B & Aya-Vision-32B & NVLM-D-72B \\
        Qwen2.5-7B & QwQ-32B & Magnum-V4-72B\\
        Mistral-7B & Qwen2.5-32B & Meta-Llama-3-70B\\
        Llama-2-7B & Aya-Expanse-32B  & Llama-3.1-70B\\
        OLMo-2-1124-7B & OLMo-2-0325-32B & Qwen2.5-72B\\
        \hline
    \end{tabular}%
    }
    \label{tab:prelim-models-evaluated}
\end{table}

\begin{table}[ht]
    \centering
    \caption{All LLMs Considered for Evaluation}
    \resizebox{0.48\textwidth}{!}{%
    \begin{tabular}{|l|l|l|}
        \hline
        \textbf{Small} & \textbf{Medium} & \textbf{Large} \\
        \hline
        
        Gemma-7B & Aya-Vision-32B & NVLM-D-72B \\
        Qwen2.5-7B & QwQ-32B & Magnum-V4-72B\\
        Mistral-7B & Qwen2.5-32B & Meta-Llama-3-70B\\
        Llama-2-7B & Aya-Expanse-32B  & Llama-3.1-70B\\
        OLMo-2-1124-7B & OLMo-2-0325-32B & Qwen2.5-72B\\
        
        GPT-Neo-2.7B & Llama-4-Maverick-17B-128E-Instruct & Llama-3.3-70B-Instruct\\
        Mistral-7B-Instruct-v0.3  & Qwen2.5-Coder-32B-Instruct & Llama-3.1-Nemotron-70B-Instruct-HF\\
        Qwen2.5-7B-Instruct & Qwen2.5-32B-Instruct &  \\
        & Deepseek-Coder-33B-Instruct&  \\
        \hline
    \end{tabular}%
    }
    \label{tab:all-models-evaluated}
\end{table}

\begin{table}[ht]
    \centering
    \caption{Chosen LLM Parameter \& Size}
    \begin{tabular}{|l|c|c|c|}
        \hline
        \textbf{LLMs} & \textbf{Parameters} &\textbf{Size}\\
        \hline
        Mistral-7B-Instruct-v0.3  & 7.3B & Small\\
        Qwen2.5-7B-Instruct & 7.61B & Small\\
        Qwen2.5-Coder-32B-Instruct & 32.5B & Medium\\
        Deepseek-Coder-33B-Instruct & 33B & Medium\\
        Llama 3.3 70B Instruct & 70B & Large\\
        Llama-3.1-Nemotron-70B-Instruct-HF & 70B & Large\\
        \hline
    \end{tabular}%
    \label{tab:llm-code-performance}
\end{table}

Our selection process considered the release date of each LLM, placing priority on newer LLMs over older LLMs. We hypothesized that newer LLMs would exhibit a higher level of efficiency and quality for code generation than older LLMs since newer LLMs are trained on larger, more comprehensive datasets~\cite{alibabacloudEvolvingLandscape}. We also focused on LLMs from trusted and well-known vendors that have elevated standards for LLM training, in the hope that this would similarly maximize the likelihood of finding the best possible LLMs for our purposes. 

Three of the six LLMs originally chosen were GPT-Neo-2.7B~\cite{huggingfaceEleutherAIgptneo27BHugging}, Mistral-7B~\cite{mistralMistralMistral}, and NVLM-D-72B~\cite{huggingfaceNvidiaNVLMD72BHugging}, all of which lacked instruction tuning. While running preliminary tests on our chosen LLMs, we discovered that these LLMs struggled to produce coherent responses when given the prompts for our generative and discriminative tasks. Because of this, we replaced all of the LLMs that were not fine-tuned to respond to instructions. 

We also encountered issues with Llama-4-Maverick-17B-128E-Instruct~\cite{huggingfaceMetallamaLlama4Maverick17B128EInstructHugging}. In the same set of preliminary tests, we struggled to run inference on the LLM with our hardware. Since this LLM has a mixture-of-experts (MoE) architecture, it requires all of its 402B parameters to be loaded into device memory simultaneously and not merely its 17B active parameters. This required significantly more CUDA memory than we had originally anticipated, so our experimental framework would have required a case-by-case inference system to have evaluated this LLM. Because the goal of our project was to fairly evaluate our chosen LLMs against one another, we could not introduce the additional independent variables that would inevitably be created by such a system.

Our final set of chosen LLMs comprised Mistral-7B-Instruct-v0.3, Qwen-2.5-7B-Instruct, Qwen2.5-Coder-32B-Instruct, Deepseek-Coder-33B-Instruct, Llama-3.3-70B-Instruct, and Llama-3.1-Nemotron-70B-Instruct-HF. Below is a brief description of each LLM: 

\begin{itemize}

\item Mistral-7B-Instruct-v0.3~\cite{huggingfaceMistralaiMistral7BInstructv03Hugging} is the fourth-best small instruction-tuned LLM according to HuggingFace user ehristoforu's collection~\cite{huggingfaceBestSmall}. This collection provides high-performing small LLMs ranging from 2B to 15B parameters. The base version of Mistral-7B~\cite{mistralMistralMistral} outperformed Llama-1-34B on several relevant benchmarks and Llama-2-13B on all applicable benchmarks. This provides evidence that smaller LLMs can potentially outperform larger LLMs, and for our dual-LLM system, this would ultimately decrease latency without sacrificing performance.
	
\item According to the Github EvalPerf Leaderboard~\cite{evalplusEvalPerfEvaluating}, Qwen2.5-7B-Instruct~\cite{Qwen257BInstructHugging} is the tenth-best LLM for efficient code generation, as well as the highest-ranking open-sourced and instruction-tuned model on the leaderboard. This chart compares the Win-Rate (WR) ratio between the LLMs' performance metrics, including their Differential Performance Score (DPS). Qwen2.5-7B-Instruct had a DPS of 84.7\%, a Pass@1 of 80, a Task WR of 52.5, and a Model WR of 72.7, performing significantly better than other "Small" open source LLMs.

\item According to the GitHub EvalPlus Leaderboard~\cite{evalplusEvalPlusLeaderboard}, Qwen2.5-Coder-32B-Instruct was marked as the highest "Medium" instruction-tuned open-source LLM on the Human Evaluation and MBPP benchmarks and had the highest overall performance. This LLM showed a significant improvement in overall code evaluation and generation from CodeQwen1.5 while matching the coding abilities of GPT-4o~\cite{huggingfaceQwenQwen25Coder32BInstructHugging}.

\item Munley et al.~\cite{munley2024llm4vv} used Deepseek-Coder-33B-Instruct~\cite{huggingfaceDeepseekaideepseekcoder33binstructHugging} to generate and analyze compiler test suites for directive-based programming models. They compared the LLM against other high-performing LLMs, including Phind-Codellama-34b-v2~\cite{huggingfacePhindPhindCodeLlama34Bv2Hugging} and GPT-4-Turbo~\cite{openaiOpenAIPlatform}. Their outcome indicated that Deepseek-Coder-33B-Instruct demonstrated the highest Pass@1 value out of all of the open-source LLMs in their selection.

\item At the time of this paper's writing, Llama-3.3-70B-Instruct~\cite{huggingfaceMetallamaLlama3370BInstructHugging} was the most up-to-date instruction-tuned Llama LLM besides Llama-4-Maverick-17B-128E-Instruct~\cite{huggingfaceMetallamaLlama4Maverick17B128EInstructHugging}. Llama-3.3-70B-Instruct was a promising candidate because it surpassed Llama-3.1-70B-Instruct on eight of its nine benchmarks, and was on par with Llama-3.1-405B-Instruct on their common benchmarks~\cite{huggingfaceMetallamaLlama3370BInstructHugging}.


\item NVLM-D-72B, the NVIDIA model we had originally selected ~\cite{huggingfaceNvidiaNVLMD72BHugging}, was inadequate for this project because it was not fine-tuned to accept instructions/prompts. We decided to replace it with Llama-3.1-Nemotron-70B-Instruct-HF~\cite{huggingfaceNvidiaLlama31Nemotron70BInstructHFHugging} so that we could evaluate at least one Nvidia LLM. Since Nvidia is such a prominent LLM vendor, we believed that collecting data on their LLMs could help contribute to other LLM research. This LLM is Nvidia's newest and most efficient instruction-tuned LLM, outperforming Llama-3.1-405B-Instruct~\cite{huggingfaceMetallamaLlama31405BInstructHugging}, Claude-3-5-Sonnet-20240620~\cite{anthropicIntroducingClaude}, and GPT-4o-2024-05-13~\cite{openaiOpenAIPlatform} on Arena Hard~\cite{githubGitHubLmarenaarenahardauto}, AlpacaEval~\cite{tatsulabAlpacaEvalLeaderboard}, MT-Bench~\cite{huggingfaceBenchHugging}, and Mean Response Length on HuggingFace.            

\end{itemize}

\subsection{Evaluation Methods} 

Since high performance on generative tasks might not be indicative of high performance on discriminative tasks (and vice versa), we chose to evaluate these two capabilities orthogonally and separated our evaluation methods into generative and discriminative categories. Our generative evaluations were designed to test the LLMs’ ability to generate directive-based parallel code that was compliant with the applicable specifications. 

Similarly, our discriminative evaluations tested the LLMs’ ability to discern whether a provided compiler test for a directive-based parallel programming model was compliant with the applicable specifications or not. In this way, we could discern which of our selected LLMs would be the best choices for our dual-LLM system.

\subsubsection{Testing Dataset}
In order to evaluate the LLMs' capabilities, we first needed a list of directive-based parallel programming model features to generate tests for, as well as a dataset of existing compiler tests with both positive and negative examples to have the LLMs judge. We collected this data by downloading existing hand-written OpenMP~\cite{githubGitHubOpenMPValidationandVerificationOpenMP_VV} and OpenACC~\cite{githubGitHubOpenACCUserGroupOpenACCVV} Validation and Verification (V\&V) Suites, and used two methods to collect the appropriate data from them:

\begin{itemize}
    \item 
    For our generative tasks, we needed a list of supported and correctly-implemented features and clauses to generate compiler tests for. To do this, we used a script to search through each hand-written compiler test in the V\&V suites and recorded every supported directive that we found, along with a list of all the clauses that were used with that directive. This allowed us to create a list of several thousand valid directive-clause combinations for our models to generate tests for, while also ensuring that all of the directive-clause combinations in the list could be compiled correctly.

    \item 
    For our discriminative tasks, we needed a collection of both valid and invalid compiler tests for the LLM to judge so that we could compare the LLM's judgments against each test's ground-truth validity. To do this, we copied the content of each compiler test from the V\&V suites and saved two copies: one copy with no alterations made, and one with a programming mistake introduced. 
    
    We used a modified version of the process described by Sollenberger et al.~\cite{sollenberger2024llm4vv}, in which they automatically introduce either a syntax error (undefined variable reference, missing closing bracket, missing semicolon), a semantic error (removing a bracketed section of code to reduce functionality, replacing the code with a randomly-generated non-directive-based serial code), or a parallel programming error (using a directive in a nonsensical manner or with inappropriate clauses, removing vital memory management routines). The only difference between their method and our method is that theirs algorithmically invalidated half of their original compiler tests, while ours created a valid and invalid copy of every test in the entire suite. This gave us a larger dataset of positive and negative examples for the LLMs to classify.
\end{itemize}

These methods allowed us to collect the data necessary for performing our evaluations and for automatically determining each LLM's performance without needing to manually review all of its responses.

\subsubsection{Generative Evaluations}

In order to evaluate each LLM's generative capabilities for directive-based parallel code, we experimented with four well-known prompting techniques to generate compiler tests for each of the features in our testing data.

\begin{itemize} 
\item \textit{Few-Shot Prompting} 
\newline 
 Few-Shot Prompting is a prompt engineering method in which multiple examples of possible responses from the LLM are provided as a part of the prompt. This provides additional context about the nature of the prompt as well as a format for the LLM to use in its own response. In our implementation, we provided several examples of valid compiler tests in our prompt to generate compiler tests for OpenMP and OpenACC across C, C++, and Fortran implementations.
\vspace{\baselineskip}

\item \textit{One-Shot Prompting}
\newline 
One-shot Prompting is very similar to Few-Shot Prompting, but provides one simple and well-crafted example as a reference for the LLM instead of a small collection of examples. This single example is meant to provide context while minimizing the possibility of the LLM over-adjusting its response to reflect the provided examples and not the instruction it has been given. In our implementation, our example also contained details about the specific features of OpenMP and OpenACC in C, C++, and Fortran. 
\vspace{\baselineskip}

\item \textit{Detailed Prompting}
\newline 
Another prompt engineering method is Detailed Prompting, where the prompt contains a large amount of detail and closely instructs the LLM on how to execute its task. This forces the LLM to follow the exact steps provided when generating its response, so a well-crafted detailed prompt can greatly improve the quality of generated code.
\vspace{\baselineskip}

\item \textit{Retrieval-Augmented Generation (RAG)}
\newline 
This assisted-generation method fetches context-appropriate information from a provided reference document to help an LLM give more accurate responses. For our project, we utilized the OpenMP and OpenACC specifications as reference documents, and used a bi-directional encoder model to identify contextually relevant information within these documents.

Our RAG system utilizes the BGE-1.5~\cite{bge_embedding} bi-directional encoder (a fine-tuned variant of BERT-large~\cite{DBLP:journals/corr/abs-1810-04805}) to generate sentence-level contextual embeddings for each sentence in the OpenMP/OpenACC specifications. These sentence-embedding pairs are then partitioned by a process we call cosine similarity clustering, which uses a reference embedding to identify the 20 most-similar sentence-embedding pairs and organizes them into a cluster. These clustered embeddings cannot be used as another cluster's reference embedding, but can still be part of additional clusters. These partitions are then represented by the average embedding of their constituent sentence-embedding pairs for more efficient similarity matching. 

During retrieval, a query is embedded using the same model, and the system first computes its similarity to all partition-level embeddings to identify the most relevant clusters. Afterwards, within these partitions, the system determines the most similar individual embeddings, which correspond to the most contextually relevant sentences from the original specifications. This two-stage retrieval process ensures both efficiency and precision in matching queries to pertinent information.
\end{itemize}

Once we generated compiler tests using these techniques, we first compiled them with an already-validated compiler. Then, we executed any files that compiled without issue and recorded their exit codes to detect runtime errors. For this paper, we considered tests that compiled and ran successfully without runtime errors to be "passing" since semantic errors would be significantly more difficult to detect automatically. Despite this potential for semantic errors in "passing" files, this generation-compilation-execution process still allowed us to rank the relative generative capabilities of our chosen LLMs by detecting syntax and functionality errors.

\subsubsection{Discriminative Evaluations}

To evaluate the discriminative capabilities of each LLM, we used two evaluation frameworks based on the LLM-as-a-Judge idea:

\begin{itemize}
\item \textit{Validation Pipeline}
\newline 
The Validation Pipeline is a framework introduced by Sollenberger et al. in~\cite{sollenberger2024llm4vv}. The method uses sequential stages to check for errors in the provided code. First, the provided code is compiled by a validated compiler. Then, it is executed to check for runtime errors. Finally, an LLM reviews the code and judges it as valid or invalid. If the code is rejected at any stage, it is not allowed to pass to the next stage of the pipeline. This framework exhibits high efficiency, both by using pipeline parallelism to evaluate code and by only allowing code that passes automated tests to be given to the LLM for review.

\item \textit{LLM-as-a-Judge Agent (LLMJ-Agent)} 
\newline 
The LLMJ-Agent is a system that integrates RAG to enhance code assessment capabilities. Unlike our generative applications of RAG, where retrieved contextual information is used to improve the quality of code generated by the LLM, our LLMJ-Agent leverages this mechanism to improve code judgment. Context-appropriate information from the specifications serve as reference material, allowing the LLM to assess code correctness more accurately. This agent-based approach allows the LLM to autonomously retrieve information that can aid in its judgments.

\end{itemize}

We also experimented with using select prompting approaches based on the approaches described by Sollenberger et al.~\cite{sollenberger2024llm4vv}. The first prompting approach (which we called the "parameter" approach) provides several parameters and criteria for the LLM to evaluate the provided compiler test against, and the second approach (which we called the "description" approach) asks the LLM to describe the functionality of the compiler test in detail and then base its judgment on that description. 

Both approaches had been shown to possess varying strengths and weaknesses by Sollenberger et al., so we wanted to see how the two approaches would compare against each other on a larger dataset and with a variety of LLMs.

The judgments made by the LLMs with each discriminative method and prompting approach were saved on a per-file basis, along with each file's ground-truth validity. This allowed us to calculate a wide variety of metrics, and also allowed us to calculate metrics later in the project that we had not originally considered when we ran our experiments.

\subsection{Metrics}

We used ten metrics to determine the best-performing LLMs. Because the two evaluation types expose different capabilities, we used separate sets of metrics for the generative and discriminative evaluation methods. The definitions and formulas for each metric are listed below.

\subsubsection{Generative Evaluation Metrics}

\begin{itemize}
    \item \textit{Compilation Rate}: The percentage of files successfully compiled from the generated test suite. This value can indicate an LLM's understanding of correct syntax, regardless of functional semantics. Ranges from 0 to 1.
\end{itemize}
\begin{equation}
    \text{Compilation Rate} = \frac{\text{\# of Files Compiled Successfully}}{\text{\# of Total Files}}
\end{equation}
\vspace{\baselineskip}

\begin{itemize}
    \item \textit{Returned-0 Rate}: The percentage of successfully compiled files that exit execution with return code 0. This can indicate an LLM's understanding of functional semantics, regardless of correct syntax. Ranges from 0 to 1.
\end{itemize}
\begin{equation}
    \text{Returned-0 Rate} = \frac{\text{\# Returned with 0}}{\text{\# of Files Compiled Successfully}}
\end{equation}
\vspace{\baselineskip}

\begin{itemize}
    \item \textit{Pass@1}: The percentage of generated files that both compiled correctly and exited with return code 0. This indicates both an understanding of correct syntax and functional semantics. Ranges from 0 to 1.
\end{itemize}
\begin{equation}
    \text{Pass@1} = \frac{\text{\# Returned with 0}}{\text{\# of Total Files}}
\end{equation}
\vspace{\baselineskip}

\subsubsection{Discriminative Evaluation Metrics}

The abbreviations TP, TN, FP, and FN are used to represent the number of True Positives, True Negatives, False Positives, and False Negatives, respectively.

\begin{itemize}
    \item \textit{Accuracy}: The percentage of correct judgments over the entire testing dataset. Ranges from 0 to 1.
\end{itemize}
\begin{equation}
    \text{Accuracy} = \frac{\textit{TP}+\textit{TN}}{\textit{TP}+\textit{TN}+\textit{FP}+\textit{FN}}
\end{equation}
\vspace{\baselineskip}

\begin{itemize}
    \item \textit{Normalized Bias}: The tendency of the LLM to prefer judgments of valid or invalid. Ranges from -1 to 1, where a normalized bias of 1 means the LLM always judges files as valid, and a normalized bias of -1 means the LLM always judges files as invalid. Determined by analyzing what kind of mistakes the LLM makes more often, when it does make a mistake; this determination then controls for unevenly-positive or unevenly-negative ground truth datasets.
\end{itemize}
\begin{equation}
    \text{Normalized Bias} = \frac{\textit{FP} - \textit{FN}}{\textit{FP}+\textit{FN}}
\end{equation}
\vspace{\baselineskip}

\begin{itemize}
    \item \textit{Permissiveness}: The percent likelihood that any mistake made by the LLM is a false positive (as opposed to a false negative). Ranges from 0 to 1.
\end{itemize}
\begin{equation}
    \text{Permissiveness} = \frac{\text{Normalized Bias}+1}{\text{2}}
\end{equation}
\vspace{\baselineskip}

\begin{itemize}
    \item \textit{Precision}: The likelihood that a judgment of valid from the LLM correctly indicates a valid file. Ranges from 0 to 1.
\end{itemize}
\begin{equation}
    \text{Precision} = \frac{\textit{TP}}{\textit{TP}+\textit{FP}}
\end{equation}
\vspace{\baselineskip}

\begin{itemize}
    \item \textit{Recall}: The percentage of valid files that the LLM was able to correctly identify. Ranges from 0 to 1.
\end{itemize}
\begin{equation}
    \text{Recall} = \frac{\textit{TP}}{\textit{TP}+\textit{FN}}
\end{equation}
\vspace{\baselineskip}

\begin{itemize}
    \item \textit{F1-Score}: The harmonic mean of the precision and recall, used as a general performance metric. Ranges from 0 to 1.
\end{itemize}
\begin{equation}
    \text{F1-Score} = \frac{2\times\text{Precision}\times\text{Recall}}{\text{Precision}+\text{Recall}}
\end{equation}
\vspace{\baselineskip}

\begin{itemize}
    \item \textit{MCC}: The Matthews Correlation Coefficient. Scores the accuracy of the LLM's classifications for both true positive and true negative examples, and often used in place of the F1-Score for this reason. Ranges from -1 to 1, with -1 indicating a complete lack of correlation, 0 indicating performance equivalent to random guessing, and 1 indicating perfect accuracy.
\end{itemize}
\begin{equation}
    \text{MCC} = \frac{(\textit{TP}\times\textit{TN})-(\textit{FP}\times\textit{FN})}{\sqrt{(\textit{TP}+\textit{FP})(\textit{TP}+\textit{FN})(\textit{TN}+\textit{FP})(\textit{TN}+\textit{FN})}}
\end{equation}

\subsection{Experimental Setup}

This experiment was conducted on Perlmutter~\cite{nerscArchitectureNERSC}, one of NERSC's HPC clusters. Each node in Perlmutter's GPU partition is equipped with an AMD EPYC 7763 CPU, 200 gigabytes of RAM, and four NVIDIA A100 GPUs with 80 GB of memory capacity each. Each GPU has a peak performance of 312 TFLOPs on BFLOAT-16s, the data type used for each LLM's parameters.


In our evaluations, our RAG system used the Beijing Academy for Artificial Intelligence's General Embedding Large English v1.5 model (BAAI's BGE-Large)~\cite{bge_embedding} to create contextual sentence-level embeddings. This is a bidirectional encoder transformer model similar in structure and functionality to BERT~\cite{DBLP:journals/corr/abs-1810-04805}. In order to maintain maximal GPU memory capacity and compute bandwidth for LLM inference, the BGE-Large model's inference was run on the CPU.

For this project, we utilized asynchronous unit experiments, with each covering a unique combination of LLM and evaluation method. With our 6 chosen models and 6 evaluation methods, 36 independent unit experiments were conducted to cover all possible combinations. 

Each experiment was launched on a single node, with a unique instance of its assigned LLM loaded onto the node's GPUs. Each node then ran its assigned evaluation on its assigned LLM and stored the results in its own record to be combined with the other nodes' results later. This setup allowed each unit experiment to cover as many testing instances as it could in its allotted time without needing to synchronize with or wait for any other unit experiment. The final evaluation was conducted on 36 nodes for 7 hours, utilizing 36 CPUs, 144 GPUs, and 7.2 TB of RAM in total.

\section{Results}
\label{sec:results}

This section contains the fine-grained and cumulative results to corroborate our findings.
We emphasized the use of the F1-Score, Pass@1, and MCC metrics to determine the best-performing models for our tasks since they can be used as single-value comprehensive indicators of general performance. Table \ref{tab:llm-code-performance} provides these metrics as calculated from all generated and evaluated files.

\begin{table}[htbp]
    \centering
    \caption{Cumulative F1-Score, MCC, \& Pass@1}
    \begin{tabular}{|l|c|c|c|}
        \hline
        \textbf{LLMs} & \textbf{F1-Score}& \textbf{MCC}&\textbf{Pass@1}\\
        \hline
        Mistral-7B-Instruct-v0.3  & .719 & .369 & .017 \\
        Qwen2.5-7B-Instruct & .723 & .390 & .077 \\
        Qwen2.5-Coder-32B-Instruct & \textbf{.735} & \textbf{.447} & .243 \\
        Deepseek-Coder-33B-Instruct & .705 & .349 & \textbf{.434}\\
        Llama-3.3-70B-Instruct & .733 & .435 & .202\\
        Llama-3.1-Nemotron-70B-Instruct-HF & .641 & .245 & .075 \\
        \hline
    \end{tabular}
    \label{tab:llm-code-performance}
\end{table}

\begin{figure}[htbp]
    \centering
    \includegraphics[width=.3\textwidth]{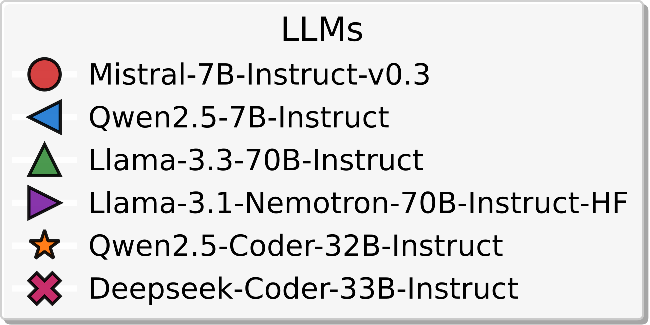}

    \includegraphics[width=.2\textwidth]{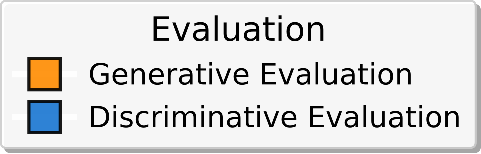}

    \includegraphics[width=.49\textwidth]{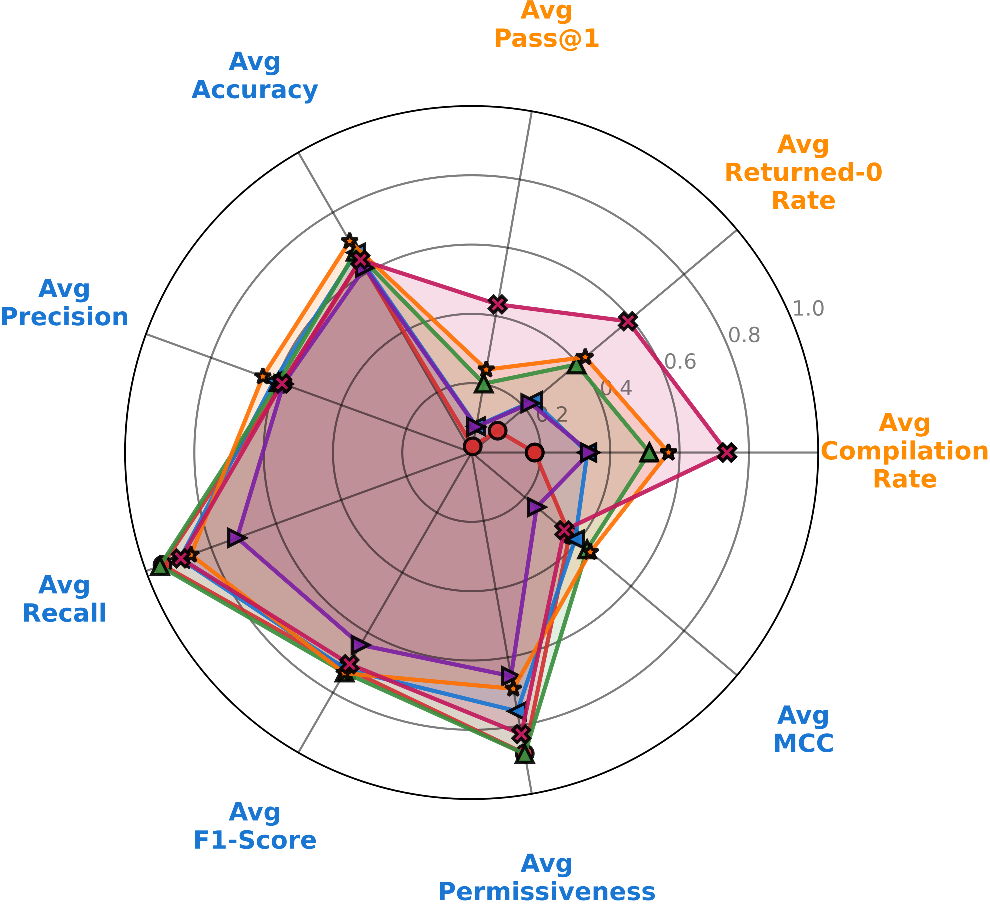}
    \caption{LLM Cumulative Performance - This Radar Chart demonstrates the cumulative performance of each LLM for all of the generative (orange) and discriminative (blue) metrics.}
    \label{fig:radar-plot}
\end{figure}

It should be noted that despite similar F1-Scores across several models, the differences in overall performance were significant. Figure \ref{fig:radar-plot} shows the Accuracy, Permissiveness, Precision, Recall, F1-Score, and MCC for all of the tested models, demonstrating wildly different strengths and weaknesses across our discriminative evaluations.

We determined each model's overall metrics by combining the results of all four generative evaluations and all two discriminative evaluations into a single set of results and calculating our values from this set. This ensured that differences in the amount of processing steps involved in each evaluation would not skew our results. This also allowed us to make a fair comparison between the LLMs, as any positive or negative factors tied to an evaluation method would equally apply to all six LLMs.

We did, however, want to know the effects that these underlying factors may have had on the scores. To better understand these factors and identify hidden trends, we reorganized our data according to the factors that could have affected it, including the programming model, programming language, evaluation method, and prompting approach that were involved in each data point's collection.

\begin{figure}[t!]
    \vspace{-17\baselineskip}
    \begin{subfigure}[t]{0.49\textwidth}
        
        \includegraphics[width=\textwidth]{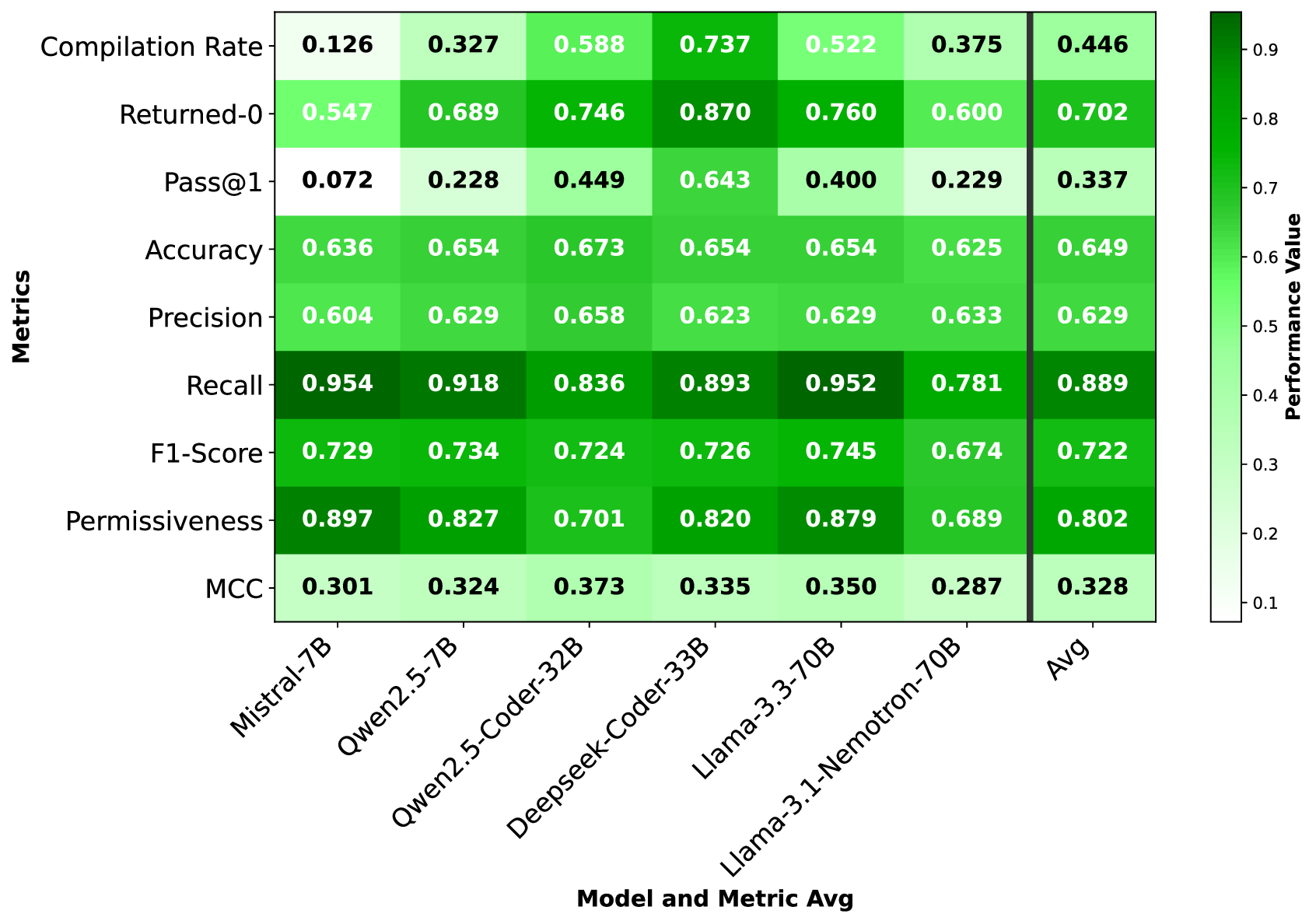}
        \caption{LLM OpenACC Performance}
        \label{fig:fine-grained-results_openacc}
    \end{subfigure}
    \begin{subfigure}[b]{0.49\textwidth}
        
        \includegraphics[width=\textwidth]{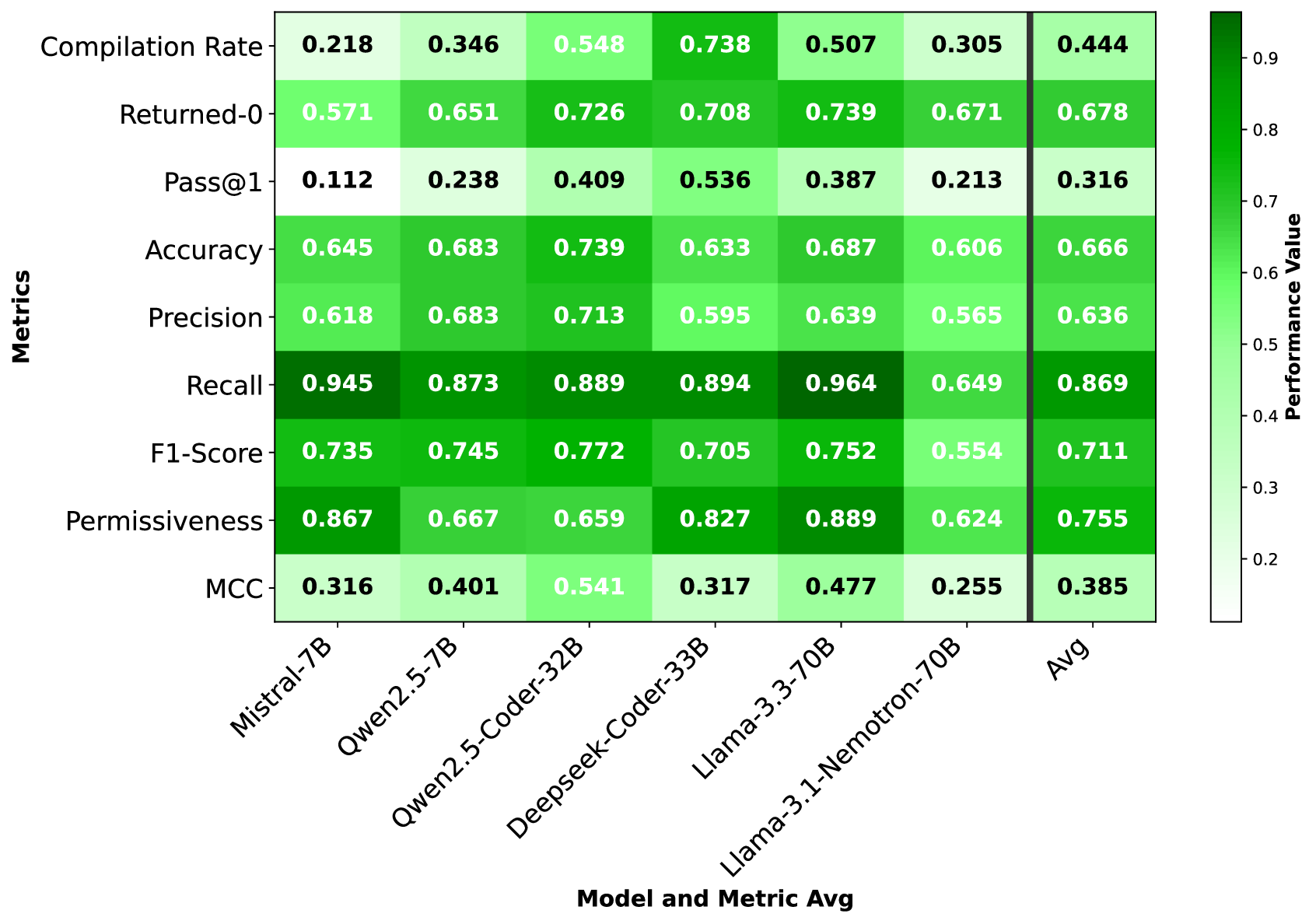}
        \caption{LLM OpenMP Performance}
        \label{fig:fine-grained-results_openmp}
    \end{subfigure}
    \caption{LLM Programming Model Performance - Correlation heatmap analysis for the fine-grained results between the two Programming Models, OpenACC and OpenMP. These heatmaps compare each metric to each model with the average of each metric displayed with a black visible line for each Programming Model.(continuation of Figure 1).}
    \label{fig:fine-grained-results_heatmaps}
\end{figure}

Figures \ref{fig:fine-grained-results_openacc} and \ref{fig:fine-grained-results_openmp} provide the metric values and their averages for OpenACC-related files and OpenMP-related files, respectively. As the average values across all six LLMs demonstrates, the LLMs performed slightly better on OpenACC generative tasks than on OpenMP generative tasks, but the discriminative results are mixed. The F1-Score was higher on OpenACC code, but the Accuracy and MCC was higher on OpenMP code. The LLMs were also generally more permissive with mistakes made on OpenACC than with mistakes on OpenMP.

\begin{figure*}[htbp]
    \centering

    \begin{subfigure}[b]{0.28\textwidth}
        \includegraphics[width=\textwidth]{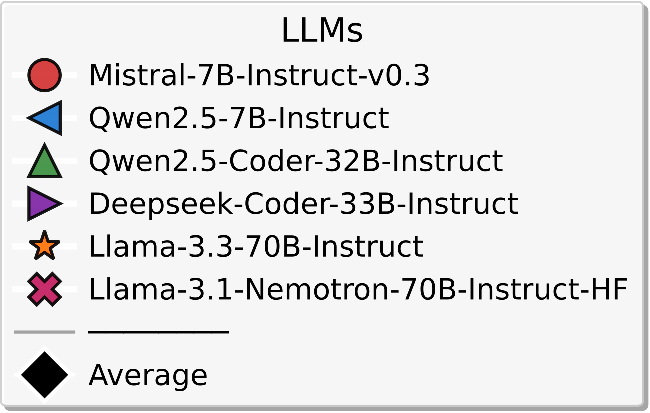}
    \end{subfigure}
    
    \begin{subfigure}[b]{0.45\textwidth}
        \includegraphics[width=\textwidth]{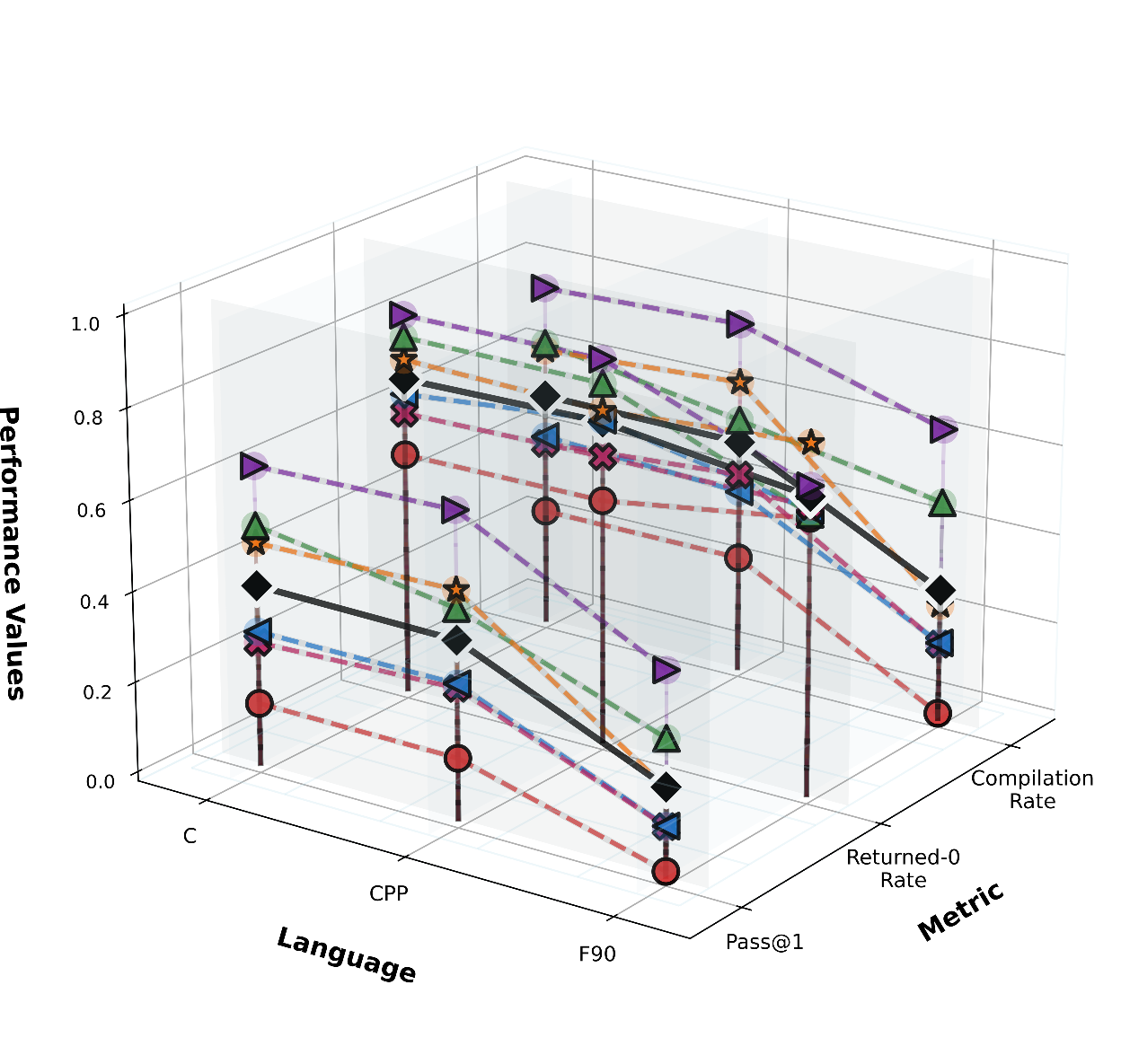}
        \caption{LLM Language Performance - The performance values for each of the LLMs based on language and metric.}
        \label{fig:fine-grained-results_language}
    \end{subfigure}
    \hfill
    \begin{subfigure}[b]{0.45\textwidth}
        \includegraphics[width=\textwidth]{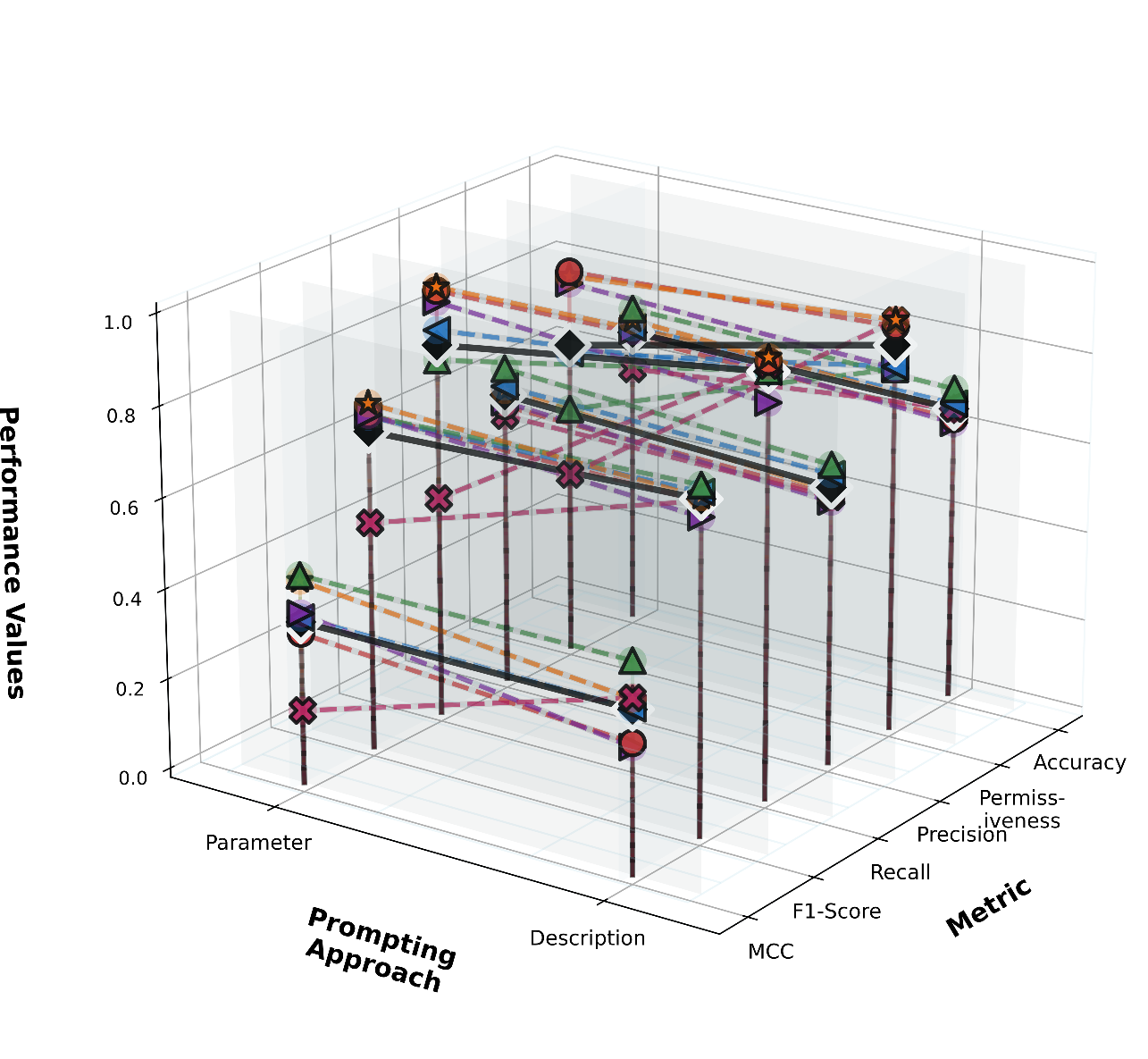}
        \caption{LLM Prompting Approach Performance - The performance values for each of the LLMs based on the prompting approach and metric.}
        \label{fig:fine-grained-results_approach}
    \end{subfigure}
    
    \vspace{0.2cm}
    
    \begin{subfigure}[b]{0.45\textwidth}
        \includegraphics[width=\textwidth]{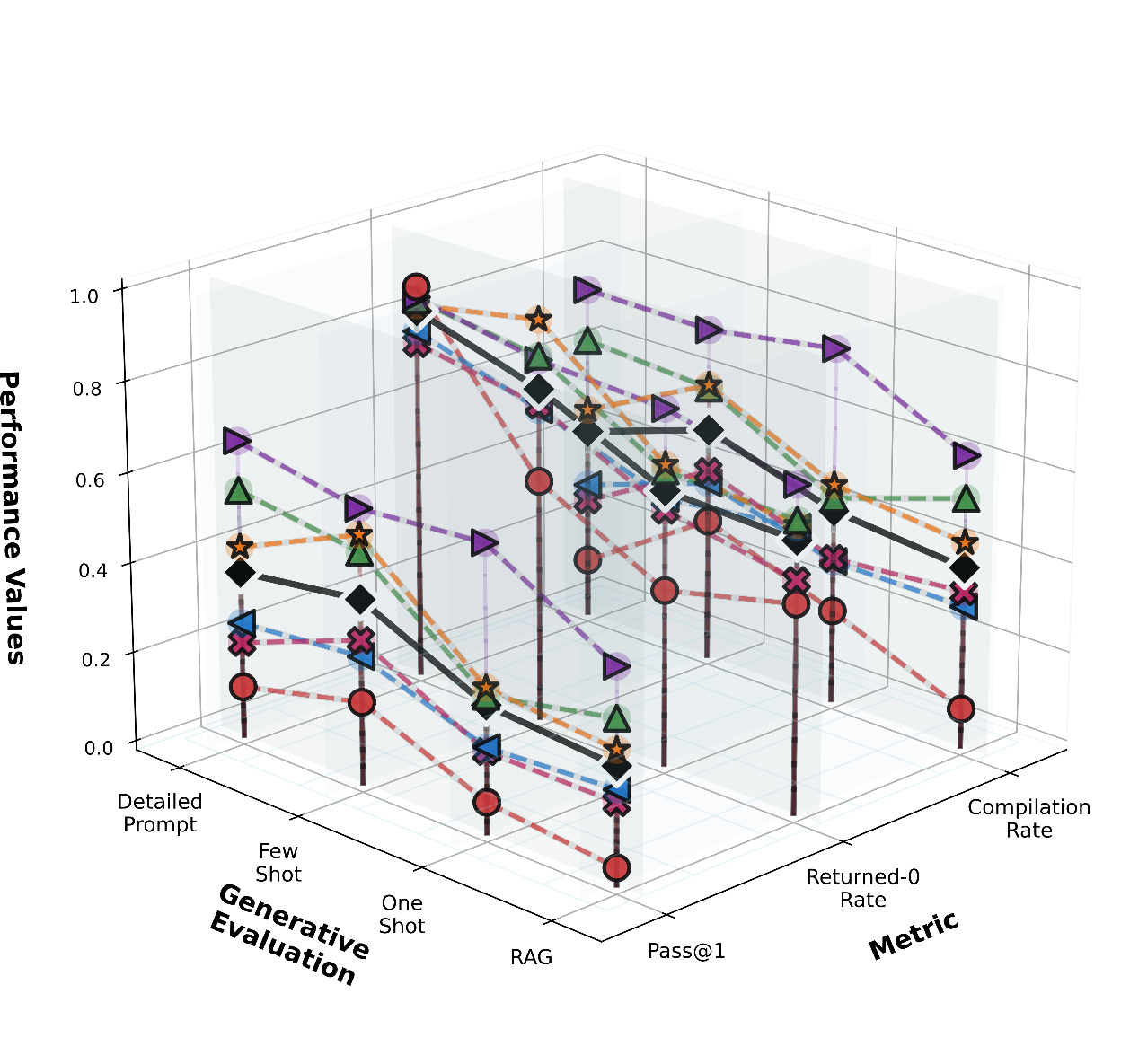}
        \caption{LLM Generative Evaluation Performance - The performance values for each of the LLMs based on the generative evaluation and metric.}
        \label{fig:fine-grained-results_gen-evals}
    \end{subfigure}
    \hfill
    \begin{subfigure}[b]{0.45\textwidth}
        \includegraphics[width=\textwidth]{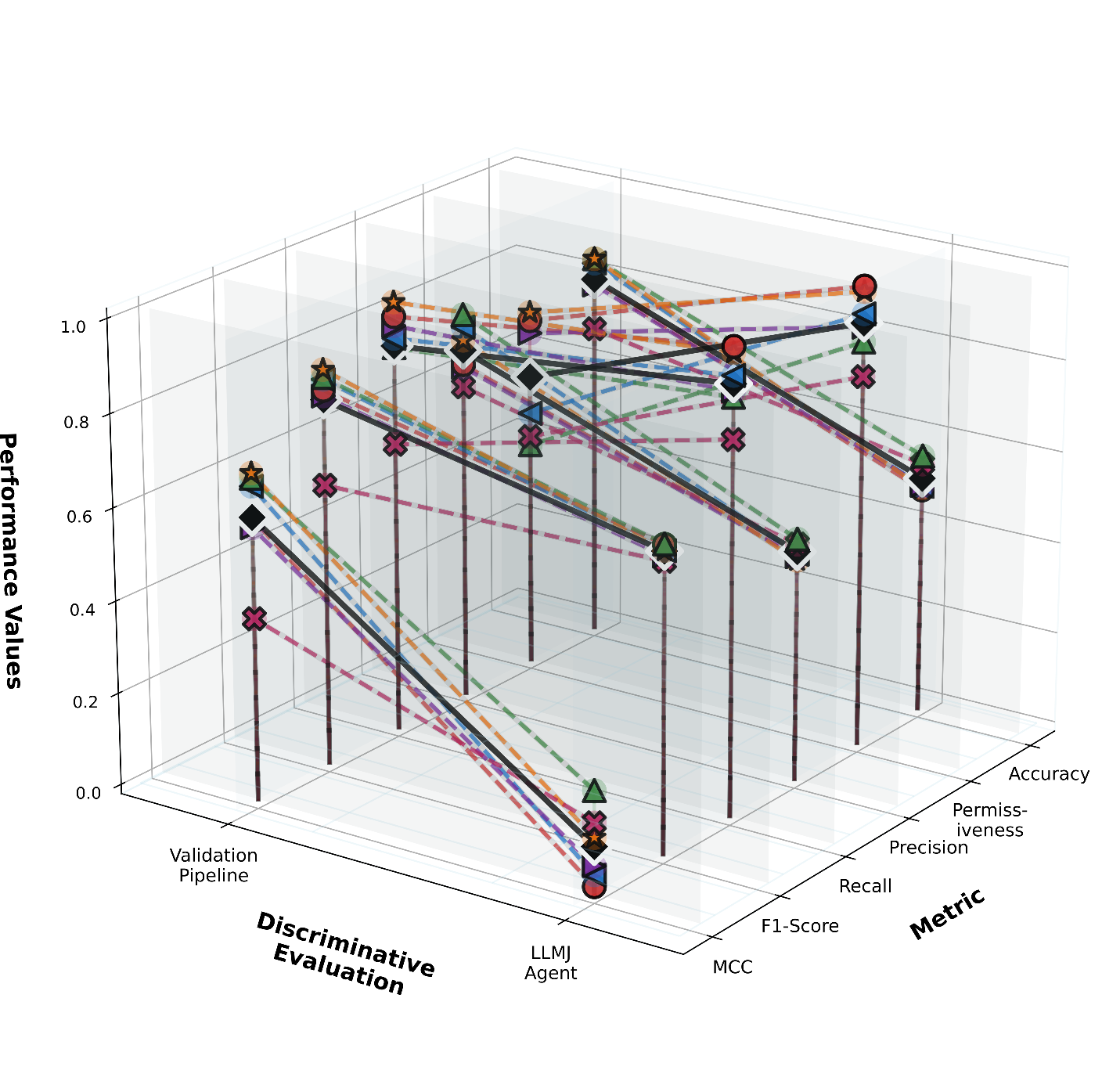}
        \caption{LLM Discriminative Evaluation Performance - The performance values for each of the LLMs based on the discriminative evaluation and metric.}
        \label{fig:fine-grained-results_disc-evals}
    \end{subfigure}
    
    \caption{This figure has four 3D fine-grained plots (a-d) that demonstrates performances of LLMs based on language and metric, based on prompting approach and metric, based on generative evaluation and metric and lastly based on discriminative evaluation and metric, respectively. One way to read the plot is for example the black diamond in these plots represents the average performance value across all six LLMs. For example, in (a) the black diamond corresponding to the F90 Language and the Pass@1 metric produces the performance value average of .193 for all six of the LLMs. As another example, in (d) the black diamond corresponding to the Validation Pipeline evaluation and MCC metric demonstrates the performance value average of .608 for all six of the LLMs.}
    \label{fig:fine-grained-results_3d-plots}
\end{figure*}

Figure \ref{fig:fine-grained-results_language} shows how LLM performances were affected by the language of the code being processed. C language had the best overall performance on both generative and discriminative evaluations, closely followed by C++. The worst overall performance was on Fortran code, likely due to an abundance of C and C++ code in modern datasets compared to the amount of Fortran code.

Figure \ref{fig:fine-grained-results_approach} shows how the prompting approaches affected the performance of the LLMs on the discriminative evaluations. On average, the Description-Based prompting approach yielded better results, though in some specific cases the Parameter-Based prompting approach was more effective. Qwen-Coder, the LLM with the best overall discriminative performance, was one such case. Because of this, we will likely use the Parameter-Based prompting approach in future works that utilize Qwen-Coder despite the Description-Based prompting approach performing better on average.

Figure \ref{fig:fine-grained-results_gen-evals} shows the performances of the LLMs on each individual generative evaluation method. On average, the Few-Shot method produced the best results, with the Detailed Prompt method as a close second.

Figure \ref{fig:fine-grained-results_disc-evals} shows the performances of the LLMs on each individual discriminative evaluation. The Validation Pipeline method showed significantly higher Accuracies, F1-Score, and MCC scores than the LLMJ-Agent, suggesting that the addition of compilation checks and execution data plays a significant role in the ability of the LLM to correctly detect problems. Based on this, we can conclude that a discriminative agent that cannot rely on compilation or execution data would have to be trained to recognize these kinds of problems explicitly, as these LLMs struggle to identify problems without this data.


Overall, the two medium-sized code-trained LLMs outperformed both the small and the large non-code-trained LLMs, with the exception of Deepseek-Coder's relatively poor discriminative performance. The Llama-3.3-70B model did perform similarly to Qwen-Coder on discriminative tasks, and had only slightly worse performance on generative tasks. Despite its large size, Llama-3.1-Nemotron demonstrated the worst relative performance on the discriminative evaluations of all of the LLMs, and relatively poor performance on generative tasks.


Across all of our generative evaluations, the highest-scoring LLM was Deepseek-Coder-33B-Instruct, far surpassing the rest of the LLMs in Compilation Rate, Returned-0 Rate, and Pass@1. Similarly, across all of our discriminative evaluations, Qwen-Coder-33B outperformed the other LLMs in terms of F1-Score, Accuracy, and Precision, though Llama-3.3-70B performed similarly and actually had a higher Recall.
Because of this, we will use Deepseek-Coder-33B-Instruct as our base LLM for generating compiler tests, and Qwen-Coder-33B as our base LLM for judging the generated tests and validating them.

\section{Conclusion and Future Work}

By comparing many open-source LLMs with parameter counts ranging up to 85 billion parameters, we were able to identify the strengths and weaknesses of each. Keeping our main goal of generating large compiler test suites in mind, we were able to determine the best LLMs to use to replace time-consuming manual creation and validation of functional and correct parallel code. These LLMs will form the core of our dual-LLM generation and validation system.

Many LLMs performed similarly in certain metrics, but varied wildly in others. Our final results showed that Deepseek-Coder-33B-Instruct exhibits the best generative capabilities out of the LLMs we tested with a Pass@1 of 0.434, while Qwen2.5-Coder-32B-Instruct exhibits the best discriminative capabilities with an F1-Score of 0.735 and an MCC of 0.447.

The next step of our project will be to fine-tune Deepseek-Coder-33B-Instruct and Qwen2.5-Coder-32B-Instruct and improve their functionality as generative and discriminative agents, respectively. Through this experiment, we were able to determine that these are the best LLMs to use for these roles, and we have taken another step towards autonomous compiler test creation for directive-based parallel programming models.

\section{Acknowledgements}
This material is based upon work supported by the U.S. DOE under Contract DE-FOA-0003177, S4PST: Next Generation Science Software Technologies Project. The authors are very grateful to OpenACC for supporting this work. This research used resources from NERSC, a U.S. DOE Office of Science User Facility located at LBNL, operated under Contract No. DE-AC02-05CH11231 using NERSC ERCAP0029463.

The authors would also like to thank Christian Munley for his guidance and contributions to this work.


\clearpage

\bibliographystyle{IEEEtran}
\bibliography{main}

\end{document}